\documentclass[letter]{aa} 

\usepackage{natbib}
\usepackage{graphicx}
\usepackage{placeins}%
\usepackage[varg]{txfonts}
\usepackage{natbib}
\usepackage{orcidlink}
\usepackage{hyperref}
\usepackage{multicol}
\hypersetup{
    colorlinks,
    citecolor=blue,
    linkcolor=blue,
    filecolor=magenta,      
    urlcolor=black,
    pdftitle={M.Birney et al.: spectro-imaging of the RU~Lupi jet and low-velocity component},
    }
\bibpunct{(}{)}{;}{a}{}{,}
\newcommand{\Msun}{M$_{\odot}$}

\newcommand{\Lsun}{L$_{\odot}$}

\newcommand{\Rsun}{R$_{\odot}$}

\newcommand{\Oi}{[O\,{\scriptsize I}]}

\newcommand{\Heib}{He\,{\scriptsize I}\,$\lambda$6678}
\newcommand{\Heic}{He\,{\scriptsize I}\,$\lambda$7065}
\newcommand{\OIa}{[O\,{\scriptsize I}]\,$\lambda$6300}

\newcommand{\CaIIa}{Ca\,{\scriptsize II}\,$\lambda$8498}
\newcommand{\CaIIb}{Ca\,{\scriptsize II}\,$\lambda$8542}
\newcommand{\CaIIc}{Ca\,{\scriptsize II}\,$\lambda$8662}

\newcommand{\Oia}{O\,{\scriptsize I}\,$\lambda$7773}
\newcommand{\Oib}{O\,{\scriptsize I}\,$\lambda$8446}
\newcommand{\SII}{[S\,{\scriptsize II}]\,$\lambda\lambda$6716,6731}
\newcommand{\SIIa}{[S\,{\scriptsize II}]\,$\lambda$6731}

\newcommand{\NII}{[N\,{\scriptsize II}]\,$\lambda$6583}

\begin{document} 
   \title{Forbidden emission line spectro-imaging of the RU~Lupi jet and low-velocity component\thanks{Based on data obtained from the ESO Science Archive Facility.}}


   \author{
       M. Birney\inst{\ref{inst1}}\orcidlink{0000-0002-6627-9863}\thanks{Corresponding author; \texttt{matthew.birney@mu.ie}},
       E.T. Whelan\inst{\ref{inst1}}\orcidlink{0000-0002-3741-9353}, 
       C. Dougados\inst{\ref{inst2}}\orcidlink{0000-0001-6660-936X}, 
       I. Pascucci\inst{\ref{inst3}}\orcidlink{0000-0001-7962-1683}, 
       A. Murphy\inst{\ref{inst4}}\orcidlink{0000-0003-0376-6127}, 
       L. Flores-Rivera\inst{\ref{inst5}}\orcidlink{0000-0001-8906-1528},
       M. Flock\inst{\ref{inst6}}\orcidlink{0000-0002-9298-3029} \\
       \and
       A. Kirwan\inst{\ref{inst7}}\orcidlink{0000-0003-3012-4509}
       \fnmsep
   }

   \institute{
       Department of Physics, Maynooth University, Maynooth, Co. Kildare, Ireland \label{inst1} 
        \and
       Univ. Grenoble Alpes, CNRS, IPAG, 38000 Grenoble, France \label{inst2}
       \and
       Lunar and Planetary Laboratory, The University of Arizona, Tucson, AZ 85721, USA \label{inst3}
       \and
       Academica Sinica, Institute of Astronomy and Astrophysics, No 1. Sec. 4, Roosevelt Rd., Taipei 10617, Taiwan, R.O.C. \label{inst4}
       \and
       Centre for Star and Planet Formation, Globe Institute, University of Copenhagen, Øster Voldgade 5-7, 1350 Copenhagen, Denmark\label{inst5}
       \and
       Max-Planck-Institut für Astronomie, Königstuhl 17, 69117 Heidelberg, Germany \label{inst6}
       \and
       School of Physics, University College Dublin, Dublin, Ireland
       \label{inst7}
   }

   \date{Received 27 September 2024 / Accepted 21 November 2024}




 
  \abstract{The first images of the jet and low-velocity component (LVC) from the strongly accreting classical T Tauri star RU~Lupi are presented. Adaptive optics-assisted spectro-imaging of forbidden emission lines was used. The main aim of the observations was to test the conclusion from a recent spectro-astrometric study that the narrow component (NC) of the LVC traces a magnetohydrodynamic (MHD) disk wind, and to estimate the mass-loss rate in the wind. The structure and morphology support a wind origin for the NC. The upper limit to the launch radius and semi-opening angle of the wind in \OIa\ emission are estimated to be 2~au and 19$^{\circ}$, in agreement with MHD wind models for high accretors. The height of the \OIa\ wind-emitting region, a key parameter for the derivation of the mass-loss rate, is estimated for the first time at $\sim$ 35~au, giving $\dot{M}_{\rm out}$ = 2.6 $\times$ 10$^{-11}$ \Msun ~yr$^{-1}$. When compared to the derived mass-accretion rate of $\dot{M}_{\rm acc}$ = 1.6 $\times$ 10$^{-7}$ \Msun ~yr$^{-1}$, the efficiency in the wind is too low for the wind to contribute significantly to the angular momentum removal.}

\titlerunning{Spectro-imaging of the RU~Lupi jet and Low Velocity Component}
\authorrunning{M. Birney et al.}

   \maketitle
%

\begin{figure}
   \centering
  \includegraphics[width=8.8cm, trim= 0cm 0cm 0cm 0cm, clip=true]{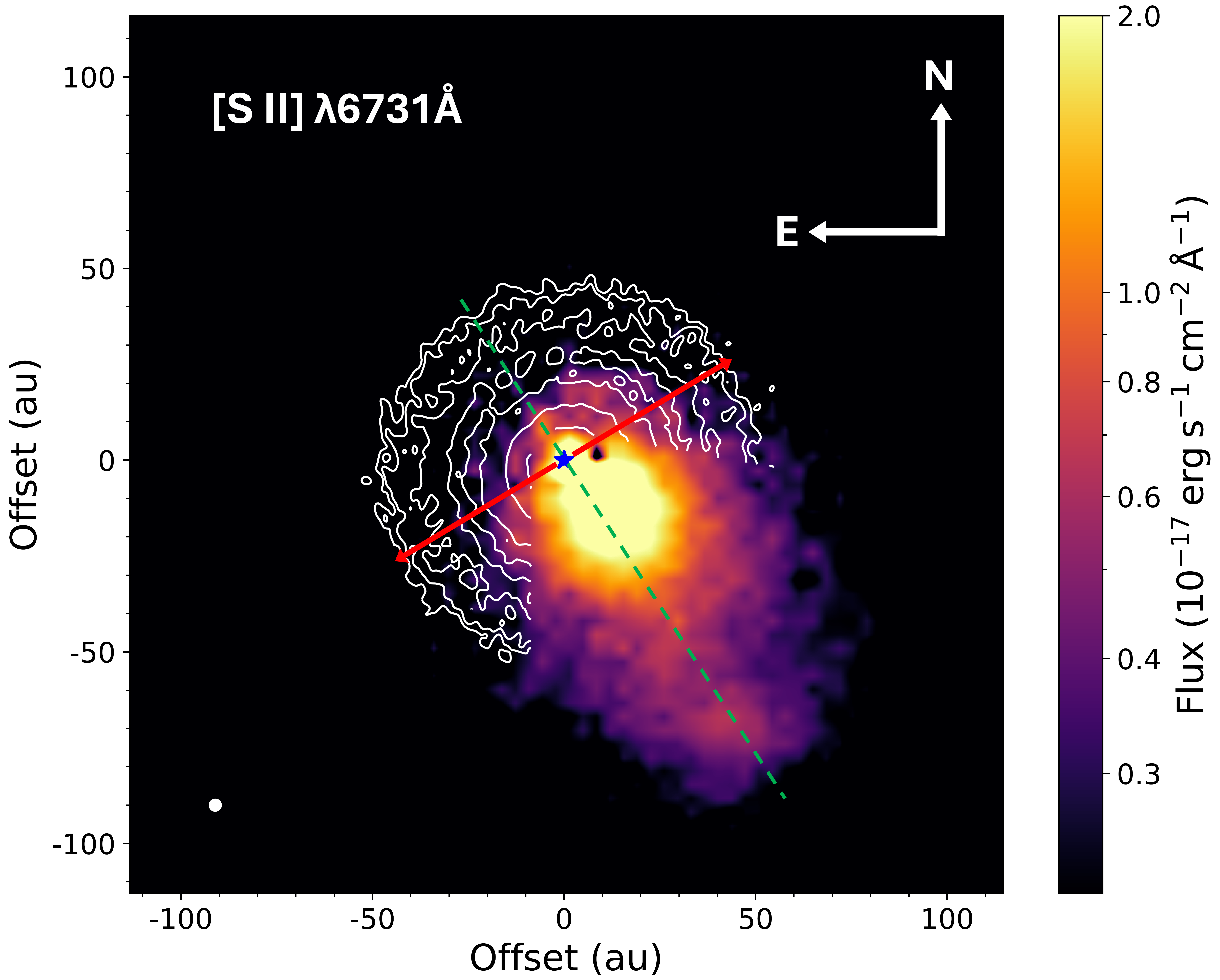}
    \caption{Integrated ($v_{\mathrm{rad}}$ = 270 -- 325 km~s$^{-1}$) \SIIa\ continuum-subtracted image of the RU~Lupi jet with a sector of the ALMA 1.3 mm dust continuum disk image shown in white contours \citep{Andrews2018}. The ALMA beam size (3.5 $\times$ 3.4 au) is shown in the lower left corner. The red arrow delineates the disk PA and radius from \cite{Huang2018} (121$^{\circ}$ $\pm$ 5$^{\circ}$, 63 $\pm$ 1 au), and the dashed green line shows the jet PA measured here (213$^{\circ}$ $\pm$ 2$^{\circ}$). The jet and outer knot are perpendicular to the disk, but the inner bright region of the jet at $<$ 50~au from the source is asymmetric about the green line.}
    \label{SII6731_image}
    \end{figure} 
    
\section{Introduction}
Classical T Tauri stars (CTTSs) are young, low-mass stars in the earliest stages of forming planets and thus are analogues of the early evolution of our Solar System \citep{Ray2007}. A key step in learning how young stars accrete mass and how their disks evolve and form planets is to clarify how angular momentum is lost \citep{PascucciPPV}. Outflows play a critical role here, as CTTSs drive outflows in the form of collimated jets and surrounding winds \citep[e.g.][]{Arulanantham2024, Bajaj2024}. Magnetohydrodynamic (MHD) winds are strongly favoured as the main mechanism by which angular momentum is removed from the accretion disk, thus regulating accretion \citep{Lesur2021, PascucciPPV}. The jet is likely the fast inner part of the MHD wind that has been collimated by the magnetic field and accelerated to velocities of $>$ 100~km~s$^{-1}$ \citep{Frank2014}. Photoevaporative (PE) winds produced by energetic radiation (UV and X-ray) from the central star heating the disk are also present, and they are critical to disk dispersal \citep{Ercolano2017}. Current work is focused on identifying direct tracers of MHD winds to test for their presence and their ability to remove angular momentum and regulate accretion. 

The low-velocity component (LVC; $v$ $<$ 40~km~s$^{-1}$) of optical forbidden emission lines (FELs) found in the spectra of CTTSs has long been a strong candidate for an MHD wind tracer \citep{Kwan1988}. As the \OIa\ line is the brightest FEL\footnote{This is due to the large abundance of oxygen and the favourable excitation conditions in the gas.} most frequently showing an LVC and very often, a high-velocity component (HVC; $v$ $>$ 40~km~s$^{-1}$), it is considered a touchstone for the LVC. The FEL HVC originates from the jet, with the emission region typically extended to several hundred au \citep{Hirth1997}. In contrast, the LVC is spatially compact, which limits what it can reveal about MHD winds \citep{Fang2023N}. A further complication is that the LVC likely also traces the PE wind \citep{Rab2023}. To uncover the origin of the LVC, studies have focused on kinematics \citep{Banzatti2019, Nisini2024N} and spectro-astrometry \citep{Whelan2021, Whelan2023}. The high velocities of the LVC coupled with emission within the potential well of the star revealed by these studies cannot easily be explained by PE winds \citep{PascucciPPV}. Adaptive optics (AO) -assisted integral field spectroscopy (IFS), such as can be done with the Multi Unit Spectroscopic Explorer (MUSE), has the capability to spatially resolve the LVC, allowing its width and height to be measured \citep{Fang2023N, Flores2023}. The width provides an upper limit of the launch radius of the outflow, and the height is required to estimate the mass-outflow rate. In combination with the mass-accretion rate, this is a direct test of its contribution to angular momentum removal \citep{Whelan2021}. Therefore, MUSE observations could potentially distinguish between the different origins for the LVC and verify whether MHD winds regulate accretion.

In this Letter, MUSE narrow-field mode (NFM) observations of the jet and wind of the CTTS RU~Lupi (d $\sim$ 140 pc) are presented. The outflow from RU~Lupi has previously been investigated with high-resolution spectroscopy and spectro-astrometry, making it an intriguing source to test which additional insights IFS can provide \citep{Takami2001, Whelan2021}. Spectro-astrometry records the spatial centroids emission as a function of velocity and therefore gives the minimum extent of the outflow, but provides no information on the width and collimation. The \OIa\ and \SIIa\ FELs of RU~Lupi were shown to have three distinct velocity components \citep{Banzatti2019}. \cite{Whelan2021} classified these components as an HVC with a peak velocity $v_{\mathrm{p}}$ $\sim$ 200~km~s$^{-1}$ (HVC2; hereafter referred to as the jet), a second, intermediate HVC that they suggested was the base of the jet (HVC1; $v_{\mathrm{p}}$ = 60 -- 100~km~s$^{-1}$) and an LVC narrow component (NC) ($v_{\mathrm{p}}$ = 13 -- 24~km~s$^{-1}$, hereafter referred to as the NC). The spectro-astrometric analysis of \cite{Whelan2021} using the Ultraviolet and Visual Echelle Spectrograph (UVES) data, revealed three new pieces of information about the RU~Lupi NC that were interpreted as it tracing an MHD wind. They were the position angle (PA) that matches the PA of the jet, an emitting region within 2~au of the star, and a negative velocity gradient with offset from the star. The MUSE observations now described offer the first images of the RU~Lupi jet and NC along with measurements of their width and height.


\section{Observations and data analysis}
RU~Lupi was observed with MUSE in NFM during July and August 2021 as part of two separate programmes (106.21EN.001 and 0104.C-0919(B)). Both sets of data were accessed through the ESO archive and produced similar results. The 106.21EN.001 data are discussed here as they were found to be of better quality due to the MUSE-DEEP data collection, where the individual observations were aligned and combined to form an observation with a higher signal-to-noise ratio (S/N)\footnote{The 0104.C-0919 (B) data are discussed further in Sect. \ref{AppSec: 0919(B) Obs}.}. MUSE covers a spectral range of 4650 -- 9300 \AA\ with a resolving power of R: 2000 -- 4000 \citep{Bacon2010}. The AO provided an angular resolution of 0\farcs074 with a spatial sampling of 0\farcs025 pixel$^{-1}$ over a field of view (FOV) of 7\farcs5 $\times$ 7\farcs5. The seeing of the observation is 0\farcs45. The data were pre-reduced using the ESO MUSE data reduction pipeline. The first step in searching for spatially resolved jet emission was to perform continuum subtraction in the jet-tracing emission lines. This was performed locally around each emission line of interest by subtracting a polynomial fit to the continuum emission from every spatial pixel in the datacube. Outflow emission was identified in the \SIIa\ line (Fig. \ref{SII6731_image}). Spatial profiles perpendicular to the jet axis were extracted at numerous positions along the jet and were fit with Gaussian profiles to measure the centroid of the spatial profiles, which are indicative of the jet axis position. In this way, the PA of the jet was accurately measured at 213$^{\circ}$ $\pm$ 2$^{\circ}$, which is perpendicular, within the uncertainty of the observations, to the estimated disk PA of 121$^{\circ}$ $\pm$ 5$^{\circ}$ \citep{Huang2018}. RU~Lupi has an almost face-on accretion disk with $\theta_{\mathrm{i}}$ = 19$^{\circ}$ $\pm$ 2$^{\circ}$ \citep{Huang2018}. The continuum-subtracted datacube was rotated so that the PA of 213$^{\circ}$ lay along the horizontal to facilitate the analysis of the extended emission (Sect. \ref{Sec: OutflowStructure}). 

   
   \begin{figure*}
   \centering
  \includegraphics[width=15.0cm, trim= 0cm 5cm 0cm 4cm, clip=true]{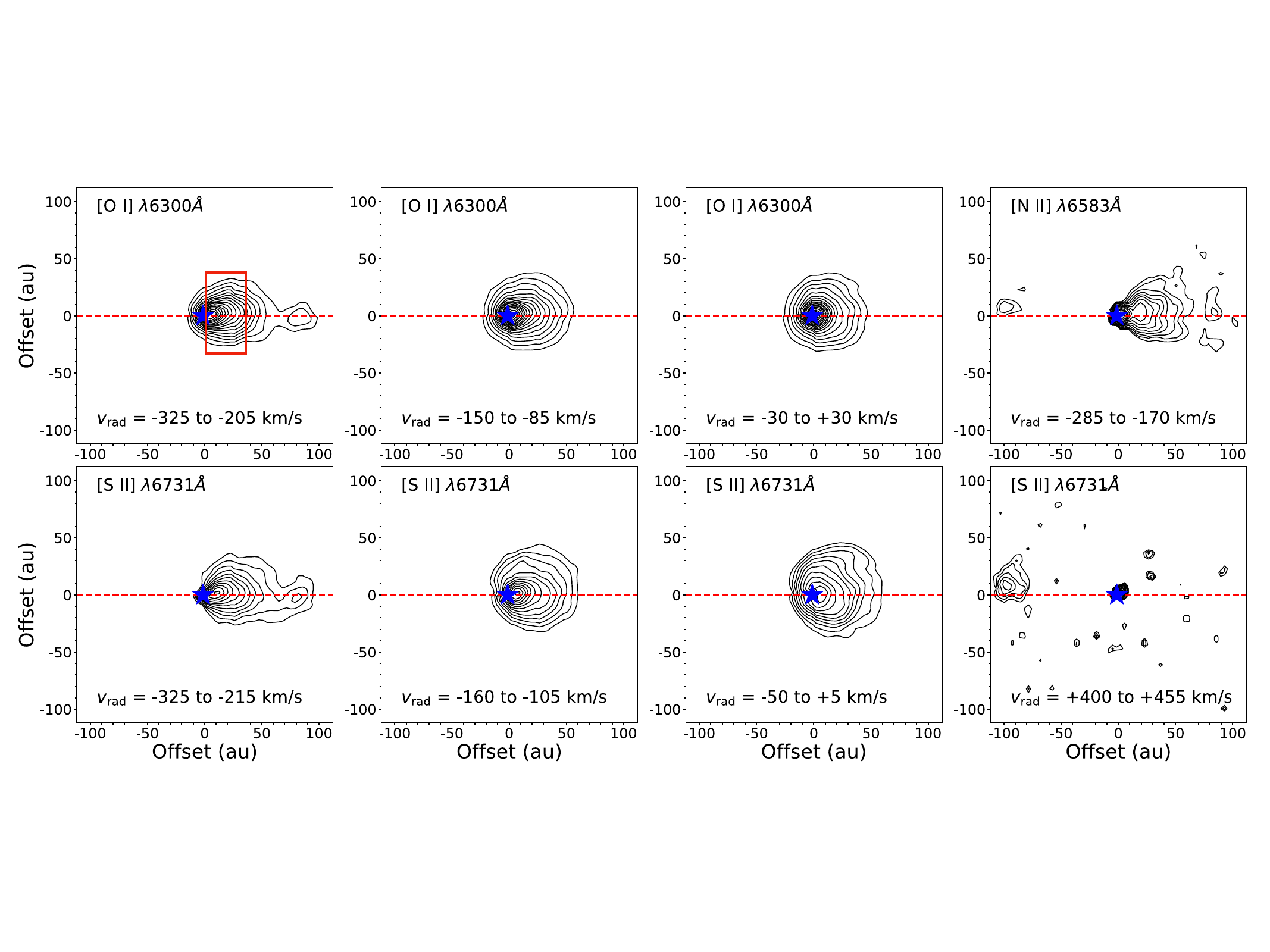}
   \caption{Deconvolved channel maps of the jet, HVC1, and NC emission from RU~Lupi in the \OIa, \SIIa\, and \NII\ emission lines. The distances are not deprojected, and the red box shows the region for which the outflow widths are mapped (Fig. \ref{collimation}). The source position is marked with a blue star. The contour levels begin at 3$\sigma$ of the background emission and increase by factors of 1.5.}
       %
       \label{deconv_channel}
    \end{figure*}

\section{Results}
The MUSE spectrum of RU~Lupi is rich in emission lines tracing accretion and outflow activity (Sect. \ref{AppSec: FullSpectrum}). The outflow is identified in the \OIa, \SII\,, and \NII\ lines, and only the jet is detected in \NII\ \citep{Hirth1997}. Faint emission at all wavelengths is found to fill the field, pointing to the presence of emission scattered from the almost face-on accretion disk. Line to continuum ratio plots were prepared to investigate the nature of the scattered emission (see Sect. \ref{AppSec: LineToContinuum}).

\subsection{Morphology of the outflow components}

Figure \ref{SII6731_image} shows an \SIIa\ image of the jet integrated over the velocity range of 270~km~s$^{-1}$ to - 325~km~s$^{-1}$. Close to the source, the outflow is wide and bright. It culminates in a knot at $\sim$ 85~au (250~au deprojected). An asymmetry across the jet axis is apparent, as the north-western side is brighter than the south-eastern side. Figure \ref{deconv_channel} presents continuum-subtracted channel maps of the jet, HVC1, and NC. A Lucy-Richardson deconvolution was performed using continuum point spread functions (PSFs) extracted from spectral regions close to the emission lines\footnote{More details on the PSFs and deconvolution are given in Sect. \ref{AppSec: ChannelsBeforeDeconv}.}. Due to the decreased spectral resolution of MUSE compared to UVES, the three outflow velocity components are not fully resolved, and any conclusions made from these channel maps should consider this. Morphological differences are present, however. In the highest-velocity channel, a small-scale jet culminating in the shock feature at $\sim$ 85~au is seen. As the \NII\ emission traces only the jet, a clear image of the jet without contamination from the other components is obtained. The morphology of the jet in \NII\ emission is wide at the base. Faint redshifted emission along the jet PA is seen in \SIIa\ at a higher velocity than in the blueshifted jet. This knot is seen at a distance where the dust disk would no longer block the redshifted flow and could be redshifted outflow emission. The velocity of this emission points to a kinematical asymmetry between the two jet lobes. The HVC1 and NC are less extended and wider than the jet, but they are extended beyond the continuum image. This suggests layering in the outflow. The asymmetry across the outflow is also visible in these images and especially in the \SIIa\ NC image.  

\subsection{Structure of the RU~Lupi outflows} \label{Sec: OutflowStructure}
Position-velocity diagrams transverse to the outflow axis were prepared from slices extracted at different distances along the outflow (Fig. \ref{PVacross}). Initially, two velocity components are present: The jet, which is seen at all distances, and the NC, which fades with increasing distance from the star. This shows that the NC is a separate outflow component to the jet (see Sect. \ref{AppSec: TransversePVlineprofiles}). To better understand the structure of the outflow, the spatial widths of the \OIa, \SIIa\, and \NII\ lines were mapped with distance from the source (Fig. \ref{collimation}). This was achieved by extracting transverse spatial profiles from the channel maps before deconvolution, and by measuring the FWHM and centroids (Fig. \ref{Fig: centroids}) using a Gaussian fitting \citep{Murphy2021, Kirwan2022AA}. Measurements were made for the inner portion of the outflow\footnote{Beyond this region, the profiles were not well fit by a Gaussian.} marked by the red box in Fig. \ref{deconv_channel}. The angular resolution was subtracted in quadrature from the measured widths to recover the intrinsic width. Measurements below the FWHM of the source PSF were excluded. Halving the lower bounds of the FWHM measurements indicates an upper limit of the radial distance along the disk from which the flows are launched (Table \ref{opening_angles}). The full opening angle of the outflow is a measure of the degree of collimation of the flow, with the jet expected to be most collimated (Table \ref{opening_angles}). For the jet, the angle between the straight line fit to the width and the y-intercept of the fit is a lower limit of the full opening angle, with the upper limit being the angle between the straight line fit through the origin. As the width of the NC is larger, especially in the case of the \SIIa\ line, the angle between the straight line fit to the width and the y-intercept likely underestimates the opening angle of the NC. We used the angle between a straight line fit through the y-intercept of the jet component as a reasonable estimate of the upper limit of the NC opening angle.

\begin{table}
\caption{Launch radii and full opening angles ($\theta$) of the jet and NC.} 
    \centering
    \begin{tabular}{cccc}
    \hline\hline  
         &Line & $r_{\mathrm{max}}$ (au) & $\theta$ ($^{\circ}$)  \\ \hline
     Jet  &\OIa &$<$ 1.0  & 24.8 -- 26.1  \\ 
        &\SIIa &$<$ 3.7& 21.0 -- 26.8  \\ 
        &\NII &$<$ 1.2 & 26.7 -- 28.4  \\
        \hline
     NC   &\OIa &$<$ 2.0 & 36.9 -- 38.4  \\ 
        &\SIIa &$<$ 9.0 & 19.9 -- 33.2  \\ 
        \hline
    \end{tabular}
    \label{opening_angles}
\end{table}

\subsection{Efficiency of the outflow components}
\begin{figure*}
   \centering
  \includegraphics[width=14.2cm, trim= 0cm 0cm 0cm 0cm, clip=true]{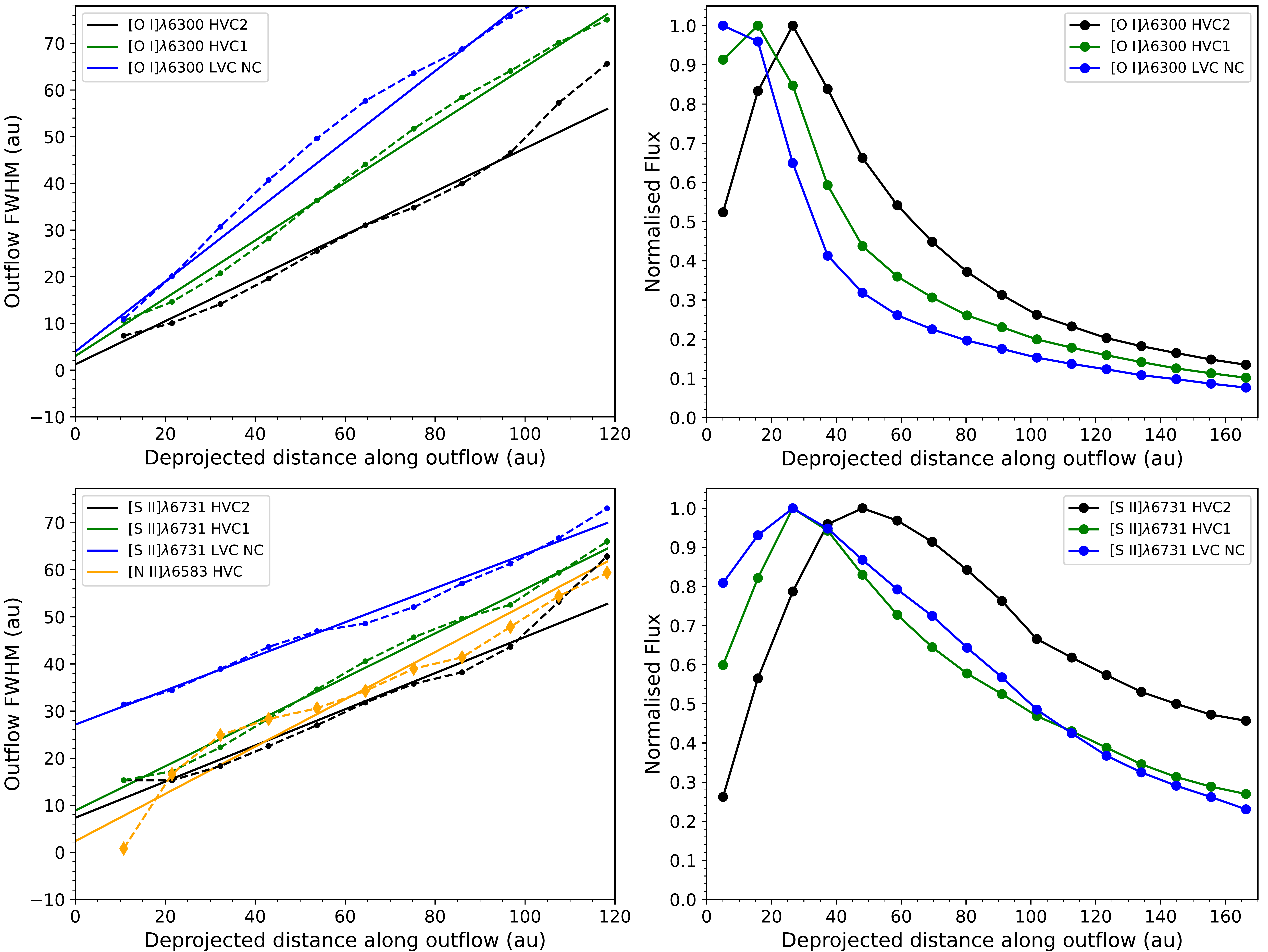}
    \caption{Full width at half maximum (left panels) and normalised flux (right panels) measurements of the jet, HVC1, and NC in \OIa, \SIIa\, and \NII\ emission, measured along the outflow, with the driving source position at the origin. The error bars on the FWHM data points are the uncertainty of the FWHM in the Gaussian fitting routine, which appear small here due to the high S/N of the transverse spatial profiles. In the right panels, the jet emission follows a different pattern than the HVC1 and NC, indicating an unresolved knot in the jet. The wind height is measured as the width of the NC emission at half the peak flux.}
    \label{collimation}
    \end{figure*}

The mass-accretion rate of RU~Lupi was estimated using the method of \cite{Alcala2017} and a mass and radius of 0.55~\Msun\ and 2.48 \Rsun. The accretion luminosities and corresponding mass-accretion rates for the accretion tracers used are given in Table \ref{Table: AccretionRates}. Some thought had to be given to the optimum aperture size for the extraction of the accretion spectrum (see Sect. \ref{AppSec: ApertureSize}). The mean value of the accretion rate measured is $\dot{M}_{\rm acc}$ = 1.6 $\times$ 10$^{-7}$ \Msun ~yr$^{-1}$. \cite{Fang2018} reported a value of $\dot{M}_{\rm acc}$ = 1.7 $\times$ 10$^{-7}$ \Msun~yr$^{-1}$ using a different set of emission lines, and \cite{Wendeborn2024} reported a value of $\sim$ 1 $\times$ 10$^{-7}$ \Msun~yr$^{-1}$ using accretion shock modelling, flow modelling, and the method of \cite{Alcala2017}, all with $L_{\mathrm{H\alpha}}$, and mitigating the effect of wind absorption, which allowed H$\alpha$ to be a more reliable accretion tracer. Using the method described by \cite{Fang2018}, the mass-outflow rate is estimated to be $\dot{M}_{\rm out}$ = 2.6 $\times$ 10$^{-11}$ \Msun ~yr$^{-1}$ for the \OIa\ NC. This method uses the velocity peak of the NC, the height of the NC emission region, and the luminosity of the NC emission. The low spectral resolution of MUSE poses a challenge here for the measurement of the NC peak velocity and luminosity. Therefore, values from \cite{Whelan2021} and \cite{Fang2018} were used in the calculation, namely $v$ = - 4~km~s$^{-1}$ and $L_{\mathrm{OI6300}}$ = 9.54 $\times$ 10$^{-6}$~\Lsun. While \cite{Whelan2021} noted variability in the jet component of the RU~Lupi outflow, the NC is seen to remain stable with time. To estimate the height of the NC, a plot of the normalised flux of the outflow components along the jet axis was produced (Fig. \ref{collimation}). The \OIa\ wind height is estimated to be 35~au from Fig. \ref{collimation}, as the width of the NC at half the height of the flux peak. This calculation assumed a temperature of 5000~K and gives a ratio of the mass outflow to accretion rate of $\sim$0.02$\%$. Taking a temperature of 10,000~K decreases this ratio by an order of magnitude, but a gas temperature of this magnitude is unlikely for the slow-moving NC. For the jet in \OIa , the method described in \cite{Fang2018} was adopted. $L_{\mathrm{OI6300}}$ = 7.43 $\times$ 10$^{-5}$~\Lsun\ was measured from the MUSE data and $L_{\mathrm{OI6300}}$ = 9.77 $\times$ 10$^{-5}$~\Lsun\ was taken from \cite{Fang2018}. The RU~Lupi HVC is known to be variable \citep{Whelan2021}. Using the luminosity from the MUSE data, a tangential jet velocity of - 52~km~s$^{-1}$, and $l_{\mathrm{\perp}}$ = 98~au, we found $\dot{M}_{\rm out}$ = 3.6 $\times$ 10$^{-9}$ \Msun ~yr$^{-1}$. This increased to $\dot{M}_{\rm out}$ = 4.7 $\times$ 10$^{-9}$ \Msun ~yr$^{-1}$ using $L_{\mathrm{OI6300}}$ from \cite{Fang2018}. The gas temperature was assumed to be 10,000 K in this calculation, as appropriate for the base of the T-Tauri jets \citep[e.g.][]{AgraAmboage2011}. The velocity estimate comes from \cite{Fang2018}, and the jet height is the projected size of the jet within the extraction aperture. The resulting outflow efficiency is in the range 2.3$\%$ to 2.9$\%$, which is expected for jets \citep{Ray2007}. 

\section{Discussion} \label{Sect: Discussion}

The full opening angle of the jet (Table \ref{opening_angles}) is large, $\sim$ 21$^{\circ}$ to 28$^{\circ}$, compared to other studies, which measured typical opening angles of $<$ 5$^{\circ}$ \citep[e.g.][]{Murphy2021}. Wide opening angles of jets have been seen in some cases, however \citep[e.g.][]{Pascucci2024}. Blending between the jet and the other outflow velocity components is ruled out as the cause, as the opening angle is also large in \NII\ emission, which only traces the jet. The jet width measurements could be impacted by the presence of a jet knot. Figure \ref{collimation} shows a peak in jet brightness at $\sim$ 25~au in \OIa\ and $\sim$ 45~au in \SIIa\, which could be an unresolved shock feature at this distance, with the difference in the peak position in \OIa\ and \SIIa\ reflecting the different critical densities the lines. The wide-angled morphology at the base of the jet may also arise from emission scattered off some V-shaped cavity carved into the surrounding ambient medium by the outflow. As the line-to-continuum ratio maps (Fig. \ref{Fig: LineToContinuum}) of the high-velocity emission also display this wide-angled morphology, we can rule out contribution from continuum light scattered off a cavity. However, we cannot dismiss the possibility that the spatial distribution of scattered jet emission differs from that of scattered continuum light, resulting in the cavity being illuminated entirely by jet emission. Conversely, the morphology of this wide-angled emission in the jet images does not match what is expected for a cavity of an almost face-on inclination \citep{Whitney2003}. 

The opening angles of the NC are smaller than what has been predicted for winds by other studies. \cite{Banzatti2019} showed that the LVC broad component (BC) and NC kinematics correlate, and that blueshifted velocities are maximised at a viewing angle of $\sim$ 35$^{\circ}$. This was taken as evidence of a conical wind geometry with a half-opening angle of 35$^{\circ}$, matching the predictions for conical-shell MHD winds made by \cite{Kurosawa2012}. However, their models, which give a half-opening angle of 35$^{\circ}$, are made for sources with a moderate mass-accretion rate (taken to be 4 $\times$ 10$^{-8}$~\Msun ~yr$^{-1}$). \cite{Kurosawa2012} showed that smaller opening angles are achieved in cases of high accretion, taken to be 3.4 $\times$ 10$^{-7}$~\Msun ~yr$^{-1}$, comparable to what we measure for RU~Lupi. The outflow displays a nested structure, with the jet, traced in \OIa\ and \SIIa, being more collimated and at a higher velocity than the NC, which is wider, moves more slowly, and is launched at larger disk radii. The NC as traced by \SIIa\ emission is wider than the \OIa\ emission, again due to the difference in the critical line densities.

Using IFS, we estimated the height of the outflow much better than it can be estimated using spectro-astrometry, which only provides a minimum offset from the source position. This allows a more accurate estimation of $\dot{M}_{\rm out}$. However, large uncertainties (Table \ref{Table: AccretionRates}) remain a challenge when estimating $\dot{M}_{\rm acc}$ and $\dot{M}_{\rm out}$ values using line luminosities. Although varying the spectral extraction aperture size when measuring the mass-accretion rate in Sect. \ref{AppSec: ApertureSize} can result in a difference of an order of magnitude, we can be relatively confident in our chosen aperture size and consequently, in our estimated $\dot{M}_{\rm acc}$, as the value agrees with values in previous studies that used different accretion tracers and different methods to estimate $\dot{M}_{\rm acc}$ \citep[e.g.][]{Fang2018, Wendeborn2024}. We would like to emphasise that the aperture size used in spectral extraction is an important factor to consider when using IFS data to estimate $\dot{M}_{\rm acc}$ and $\dot{M}_{\rm out}$.

\section{Summary and conclusions}

The MUSE NFM observations presented here provide new insights into the RU~Lupi jet and NC. They build on what was previously learned from spectro-astrometry and support the conclusion that the \SIIa\ and \OIa\ NC traces an MHD wind. Our main conclusions are summarised as follows.

\begin{enumerate}

\item The jet was imaged for the first time in the \OIa , \SIIa , and \NII\ forbidden emission lines. It displays a wide full-opening angle ( $\sim$ 21$^{\circ}$ to 28$^{\circ}$) and culminates in a knot at $\sim$ 85 au (250 au deprojected), with a spectacular spatial resolution of 0\farcs74. The opening angle of the jet is large compared to other studies (typically $<$ 5$^{\circ}$), but the jet width measurements could be impacted by the presence of an unresolved knot in the inner outflow region or by jet emission scattered off a V-shaped cavity. It is possible that the jet possesses a kinematical asymmetry when the high-velocity redshifted emission detected in \SIIa\ is considered. 

\item The jet extends to double the minimum distance suggested by the spectro-astrometric analysis, and its PA (213$^{\circ}$ $\pm$ 2$^{\circ}$) is perpendicular to the RU~Lupi disk PA (121$^{\circ}$ $\pm$ 5$^{\circ}$) within the uncertainties. 

\item The outflow efficiency of the jet is at a few percent, which is in line with the values derived for other CTTSs \citep{Ray2007}. The outflow efficiency in the \OIa\ NC is very low, even when all the uncertainties in the calculations are considered, such as poorly understood wind geometries, spatial extents, and contribution to emission from the disk surface \citep{Nisini2024N}. The mass-loss rate value places the efficiency of the outflow component that is traced by the NC at $<$ 0.1$\%$, while models suggest that $\dot{M}_{\rm out}$ $\sim$ $\dot{M}_{\rm acc}$ for the wind if it dominates angular momentum removal \citep{Bai2017}. The measured mass-loss rate is also relatively low for PE winds \citep{PascucciPPV}.

\item The opening angles for the NC are smaller than what has been predicted for winds by other studies ($\sim$ 70$^{\circ}$). However, narrower opening angles are predicted in cases of high accretors such as RU~Lupi. 

\item The NC was also imaged for the first time, allowing its height in \OIa\ to be estimated. The value of $\sim$ 35~au measured from Fig. \ref{collimation} is higher than the 8~au from the \OIa\ spectro-astrometry, which provides a minimum offset. The NC is wider than the jet in \OIa\ and \SIIa, with the \SIIa\ emission being wider than the \OIa\ emission due to the difference in the critical line densities. This nested velocity structure supports the wind origin for the NC. 

\end{enumerate}



The key result is that the mass-loss rate is significantly lower in the \OIa\ NC than in the jet. \cite{PascucciPPV} argued that with high accretors such as RU~Lupi, the wind is predominantly molecular, meaning that atomic wind tracers such as \Oi\ do not trace the bulk of the wind mass-outflow rate and instead provide a lower limit on $\dot{M}_{\rm out}$. For RU~Lupi, \cite{Pontoppidan2011} reported that the CO fundamental line at $\sim$ 4.7$\mu$m traces a wind, and they estimated $\dot{M}_{\rm out}$ of the CO wind at 4 $\times$ 10$^{-9}$ \Msun ~yr$^{-1}$, which is two orders of magnitude higher than the \OIa\ NC. This supports the explanation for the low \OIa\ NC mass-loss rate offered by \cite{PascucciPPV}, but the efficiency of the CO outflow is comparable to the FEL jet ($\sim$ 2$\%$), which means that the CO wind does not dominate angular momentum removal either. This work strongly highlighted the power of IFS for advancing our understanding of the best MHD wind tracer and whether they should be considered important to angular momentum removal in forming stars. VLT/ERIS observations of RU~Lupi would be a logical next step as the molecular wind in H$_2$ emission could be analysed in the same way as was done here for \OIa\ and with an improved spectral resolution.

\begin{acknowledgements}
Work supported by the John \& Pat Hume Doctoral Scholarship at Maynooth University (MU) and the Travelling Doctoral Studentship from the National University of Ireland (NUI). We would also like to thank Carlo Manara and Andrea Banzatti for their insightful discussions. A. M. acknowledges the grant from the National Science and Technology Council (NSTC) of Taiwan 112-2112-M-001-031- and 113-2112-M-001-009-. L. Flores-Rivera acknowledges support from the ERC Starting Grant 101041466-EXODOSS. 
\end{acknowledgements}

\bibliography{Bibliography}
\bibliographystyle{aa}

\clearpage

\begin{appendix}

\section{Full Spectrum and MUSE line profiles} \label{AppSec: FullSpectrum}
\begin{figure}[h!]
   \centering
  \includegraphics[width=7.8cm, trim= 2cm 0cm 0cm 0cm, clip=False]{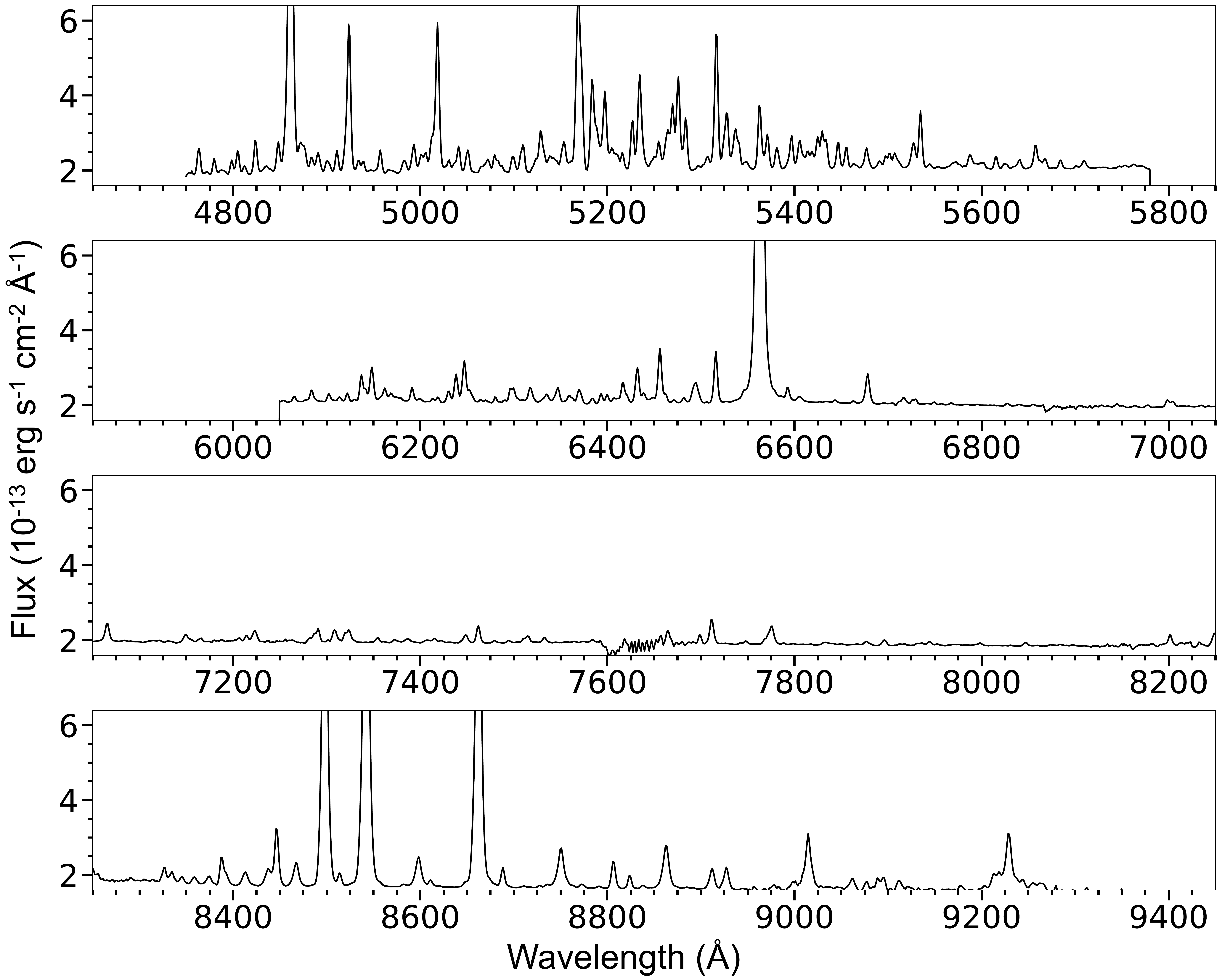}
    \caption{The full MUSE spectrum of RU~Lupi extracted at the source position with an aperture of radius 98~au.}
    \label{Fig: spec_full}
    \end{figure}

\begin{figure}[h!]
   \centering
  \includegraphics[width=7.3cm]{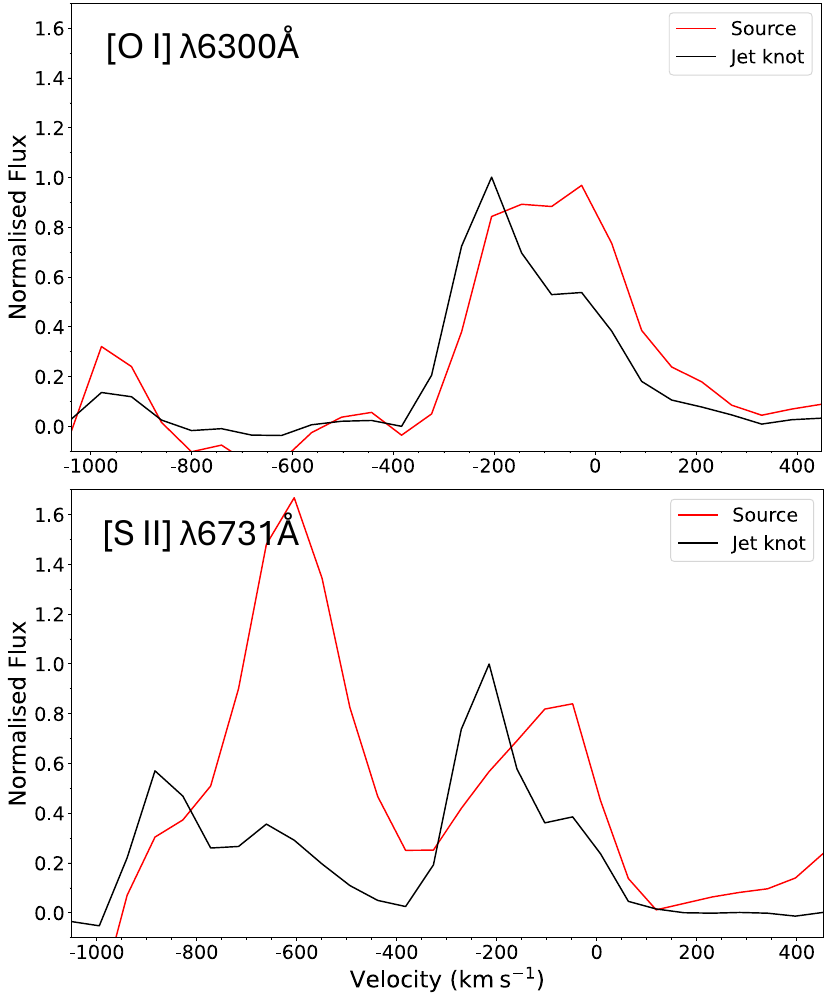}
    \caption{The MUSE \OIa\ and \SIIa\ emission line profiles. The red profiles display the spectra extracted using a circular aperture of radius 14~au from the source position. The black profiles display the spectra extracted using a circular aperture of radius 14~au from the jet knot position located at approximately 85~au along the jet.}
    \label{Fig: spec_comp}
    \end{figure}

The full MUSE spectrum extracted at the source position with an aperture of radius 98~au is presented in Fig. \ref{Fig: spec_full}. See Sect. \ref{AppSec: ApertureSize} for the choice of aperture size. There is a gap in the spectral range from 5780 -- 6050 \AA\ due to the sodium laser AO system. There is a telluric absorption feature at 7600 \AA\ due to molecular oxygen (O$_2$). Figure \ref{Fig: spec_comp} shows the \OIa\ and \SIIa\ line profiles extracted from the source position and from the jet knot at 85~au. Two velocity components are visible, however they are not well resolved due to the low spectral resolution of MUSE and the HVC1 is not detected by eye. For this reason the radial velocity measurements used in this study are taken from the UVES data of \cite{Whelan2021}.

\FloatBarrier
\section{Line to continuum ratio maps} \label{AppSec: LineToContinuum}

Line to continuum ratio plots are an effective tool for investigating the contribution of scattered emission to intrinsic line emission \citep{Delabrosse2024, Birney2024}. These figures display the result of calculating the ratio between an integrated line image, before continuum subtraction, and an integrated continuum image that is taken spectrally nearby to the line of interest. An assumption is made that the scattered emission is entirely light from the driving source of the outflow scattered off the disk and surrounding nebulous material. Therefore, the contribution from scattered light to the integrated line images will be represented by a constant ratio in these plots, as the scattered light follows the same spatial distribution as the continuum emission. In contrast, the real, intrinsic local emission will appear bright in these images, exceeding the constant ratio that represents the contribution from scattered light. 

\begin{figure*}
   \centering
    \includegraphics[width=18cm, trim= 0cm 0cm 0cm 0cm, clip=true]{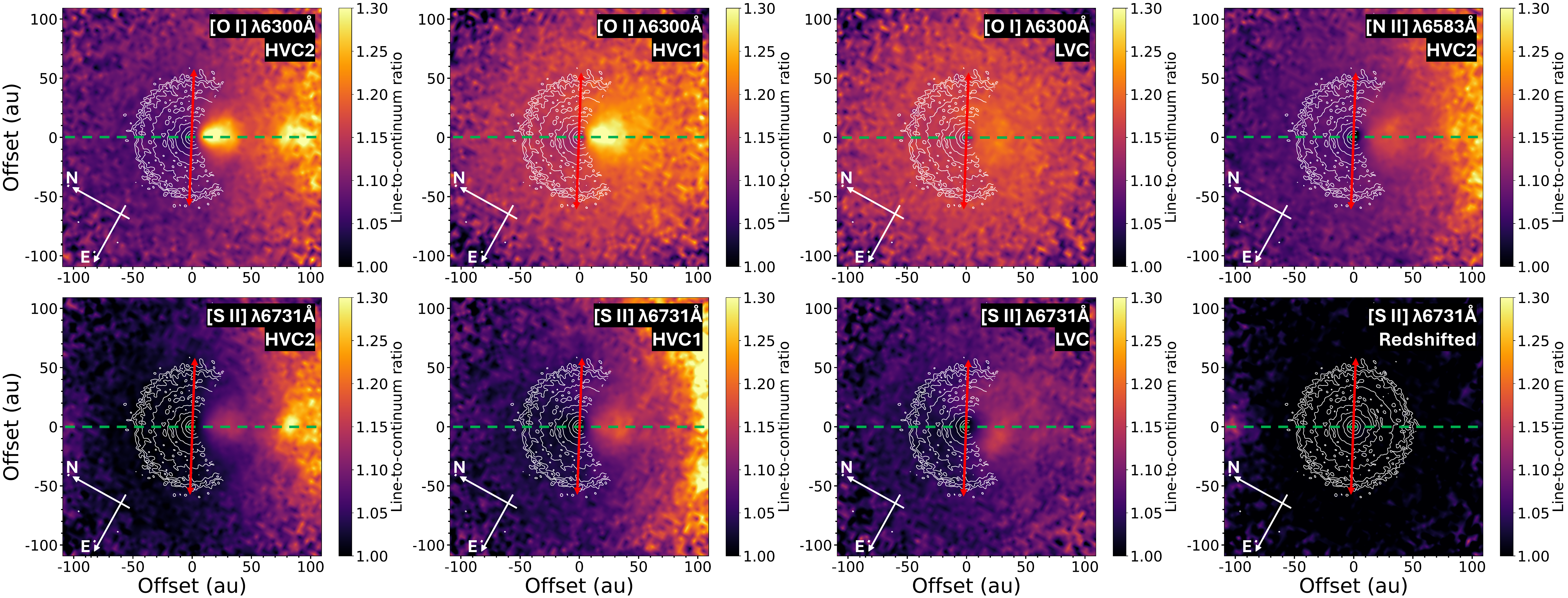}
   \caption{Line to continuum ratio maps of the \OIa, \SIIa$\,$and \NII$\,$channel maps shown in Figs. \ref{Fig: OIchannels_compare} and \ref{Fig: SIIchannels_compare}, before continuum subtraction and deconvolution. A sector of the ALMA 1.3 mm dust continuum disk image is shown in white contours. Intrinsic line emission appears bright in these images, whereas scattered light is represented by a constant ratio close to unity. The red arrow delineates the disk PA and radius from \cite{Huang2018} (121$^{\circ}$ $\pm$ 5$^{\circ}$, 63 $\pm$ 1 au) and the green dashed line represents the jet PA measured here (213$^{\circ}$ $\pm$ 2$^{\circ}$).}
    \label{Fig: LineToContinuum}
\end{figure*}

Figure \ref{Fig: LineToContinuum} displays the line to continuum ratio plots for the \OIa, \SIIa$\,$and \NII$\,$channel maps shown in Figs. \ref{deconv_channel}, before continuum subtraction and deconvolution (see Sect. \ref{AppSec: ChannelsBeforeDeconv} for the channel maps before and after deconvolution). A sector of the ALMA 1.3~mm dust continuum disk image is shown in white contours \citep{Andrews2018}. In the high-velocity channels (HVC2) one sees the intrinsic collimated jet emission extending from the source position and culminating in a jet knot at an offset from the source of approximately 85~au. In the intermediate velocity channel (HVC1) one also sees the base of the outflow in intrinsic line emission, but comparatively less collimated than the jet emission seen in the HVC2 line to continuum ratio maps.

Also present in the jet and HVC1 line to continuum ratio maps is a wide, arc-like feature, located just beyond the jet knot at 85~au seen in the HVC2 channels. This arc-like feature is not detected in the channel maps shown in Figs. \ref{Fig: OIchannels_compare} and \ref{Fig: SIIchannels_compare}. Intrinsic emission in line to continuum ratio maps appears brighter as offset from the source increases as the contribution from scattered light decreases with distance from the source. As this wide arc-like feature is not detected in the channel maps, it is likely that it is intrinsic, very low-intensity line emission that appears bright in the line to continuum maps, as the contribution from scattered light is very low in the region, resulting in a large line to continuum ratio. As the feature displays an arc-like morphology, it may indicate the location of the edge of the gas disk surrounding RU~Lupi, where the inner edge of the arc-like feature is the outer edge of the gas disk. Small dust particles responsible for scattering at optical wavelengths follow the spatial distribution of the gas disk, that can extend two to three times further than mm dust disk. As the contribution from scattered light is likely very low in the region where the the arc-like feature is located, this indicates a rapid decrease in the density of scattering particles which would coincide with the gas disk edge. 

In the low-velocity channels (LVC) one sees intrinsic wide-angled outflow emission, that is comparatively dimmer than the jet emission. In the high-velocity redshifted \SIIa$\,$emission, labelled Redshifted in Fig. \ref{Fig: LineToContinuum} one sees an intrinsic knot of line emission at an offset of approximately 100~au from the source position, in the opposite direction of the blueshifted outflow. In the HVC1 and LVC channels in \OIa\ emission, one sees significant line emission across the FOV. We attribute this emission to residual scattered line emission from the outflow, that does not follow the same spatial distribution as scattered light from the source, resulting in the outflow emission illuminating surrounding ambient material.

\clearpage
\onecolumn

\begin{multicols}{2}
\section{Results before and after deconvolution}\label{AppSec: ChannelsBeforeDeconv}
\begin{figure*}[h!]
\centering
\includegraphics[width=17cm, trim= 0cm 0cm 0cm 0cm, clip=true]{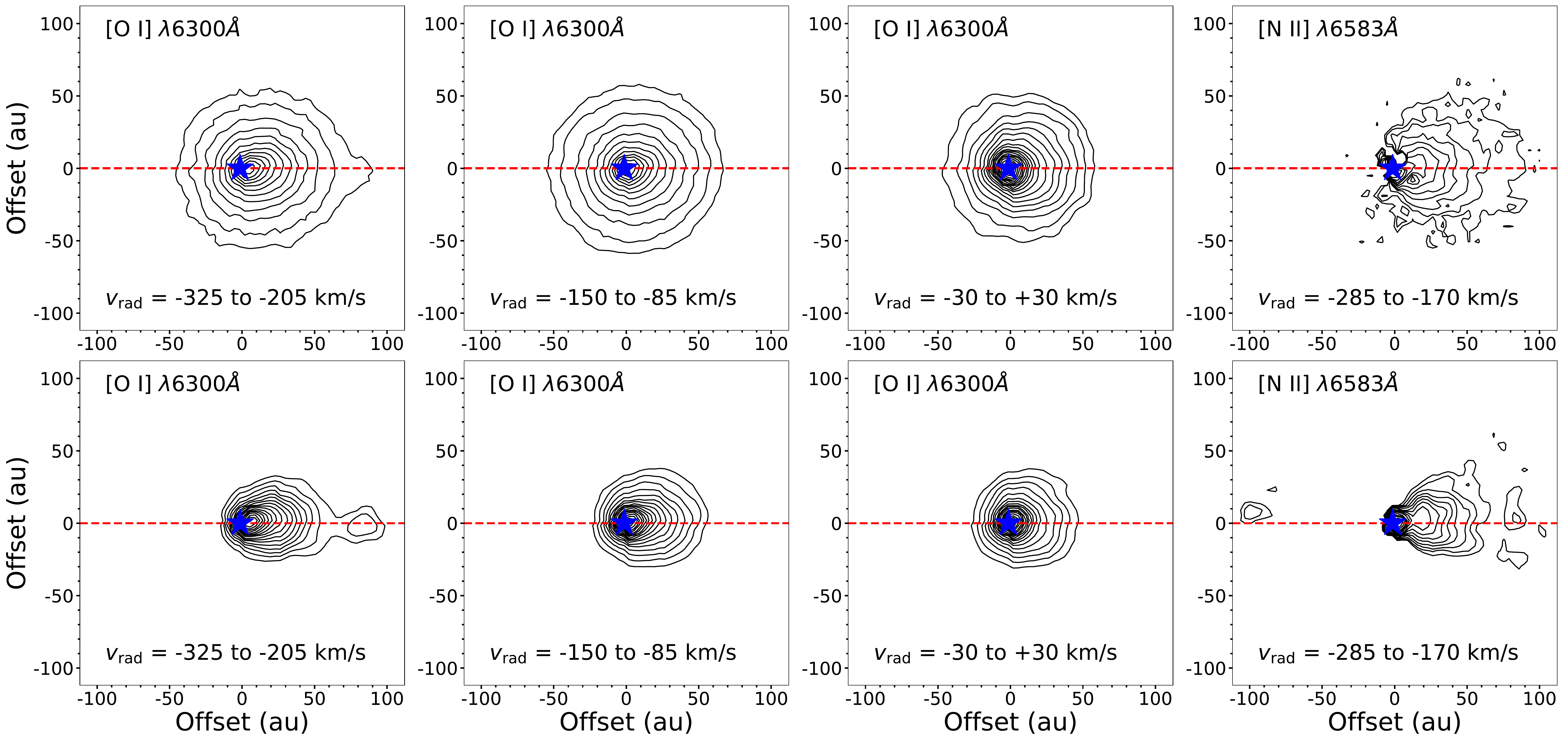}
   \caption{Comparing the channel maps of the jet, HVC1 and NC in \OIa\ emission, and the jet in \NII\ emission, before (top panels) and after (bottom panels) deconvolution. The source position is marked with a blue star. The contour levels begin at 3$\sigma$ of the background emission and increase by factors of 1.5.}
   \onecolumn
       \label{Fig: OIchannels_compare}
\end{figure*}

\begin{figure*}[h!]
   \centering
    \includegraphics[width=17cm, trim= 0cm 0cm 0cm 0cm, clip=true]{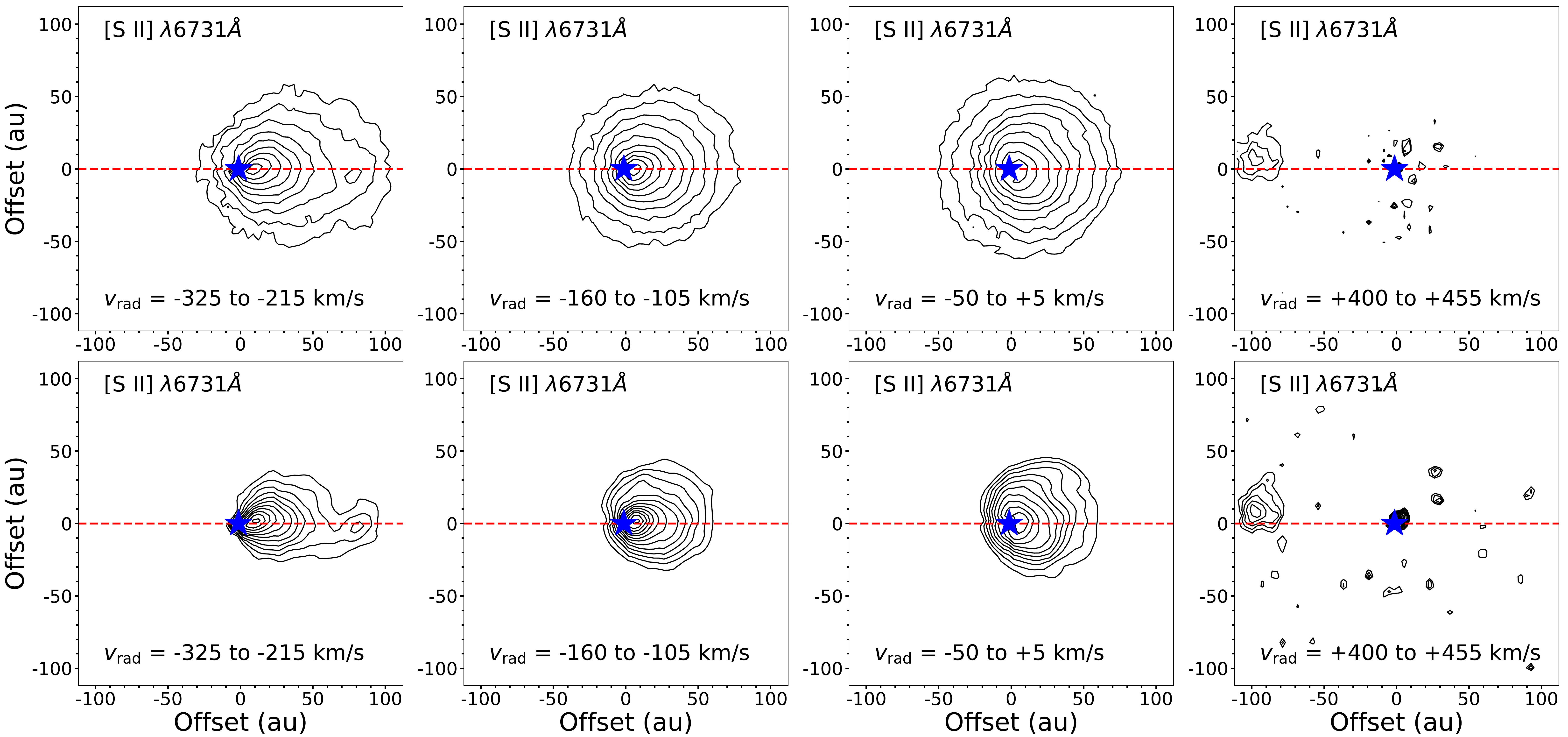}
   \caption{Comparing the channel maps of the jet, HVC1 and NC in \SIIa\ emission, and the high-velocity redshifted knot, also in \SIIa\ emission, before (top panels) and after (bottom panels) deconvolution. The source position is marked with a blue star. The contour levels begin at 3$\sigma$ of the background emission and increase by factors of 1.5.}
   \onecolumn
       \label{Fig: SIIchannels_compare}
    \end{figure*}

\begin{figure*}[h!]
   \centering
    \includegraphics[width=14cm, trim= 0cm 0cm 0cm 0cm, clip=true]{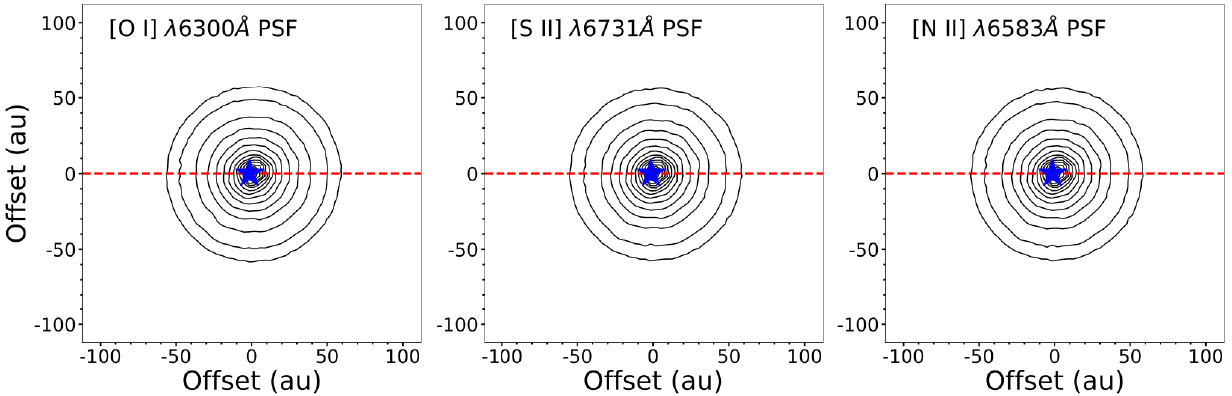}
   \caption{The PSFs used when deconvolving the \OIa\ , \SIIa\ , and \NII\ emission. The source position is marked with a blue star. The contour levels begin at 3$\sigma$ of the background emission and increase by factors of 1.5. }
       \label{Fig: PSFfigure}
    \end{figure*}

Figures \ref{Fig: OIchannels_compare} and \ref{Fig: SIIchannels_compare} display the channel maps before (top panels) and after (bottom panels) the Lucy-Richardson deconvolution is applied. An assumption was made when estimating the PSF used for the deconvolution, in that the local continuum emission was used as the PSF. The PSFs used when deconvolving the \OIa, \SIIa, and \NII\ emission are shown in Fig. \ref{Fig: PSFfigure}. The deconvolution was allowed to run for 20 iterations to enhance the morphology of the outflow without introducing any artefacting. The jet knot at $\sim$ 85~au is marginally visible before deconvolution in the high-velocity channels. The possible high-velocity \SIIa\ redshifted knot is also visible before deconvolution. 
\end{multicols}

\FloatBarrier
\begin{multicols}{2}
\section{Transverse PV diagrams and line profiles}\label{AppSec: TransversePVlineprofiles}
\begin{figure*}
   \centering
  \includegraphics[width=18cm, trim= 0cm 0cm 0cm 0cm, clip=true]{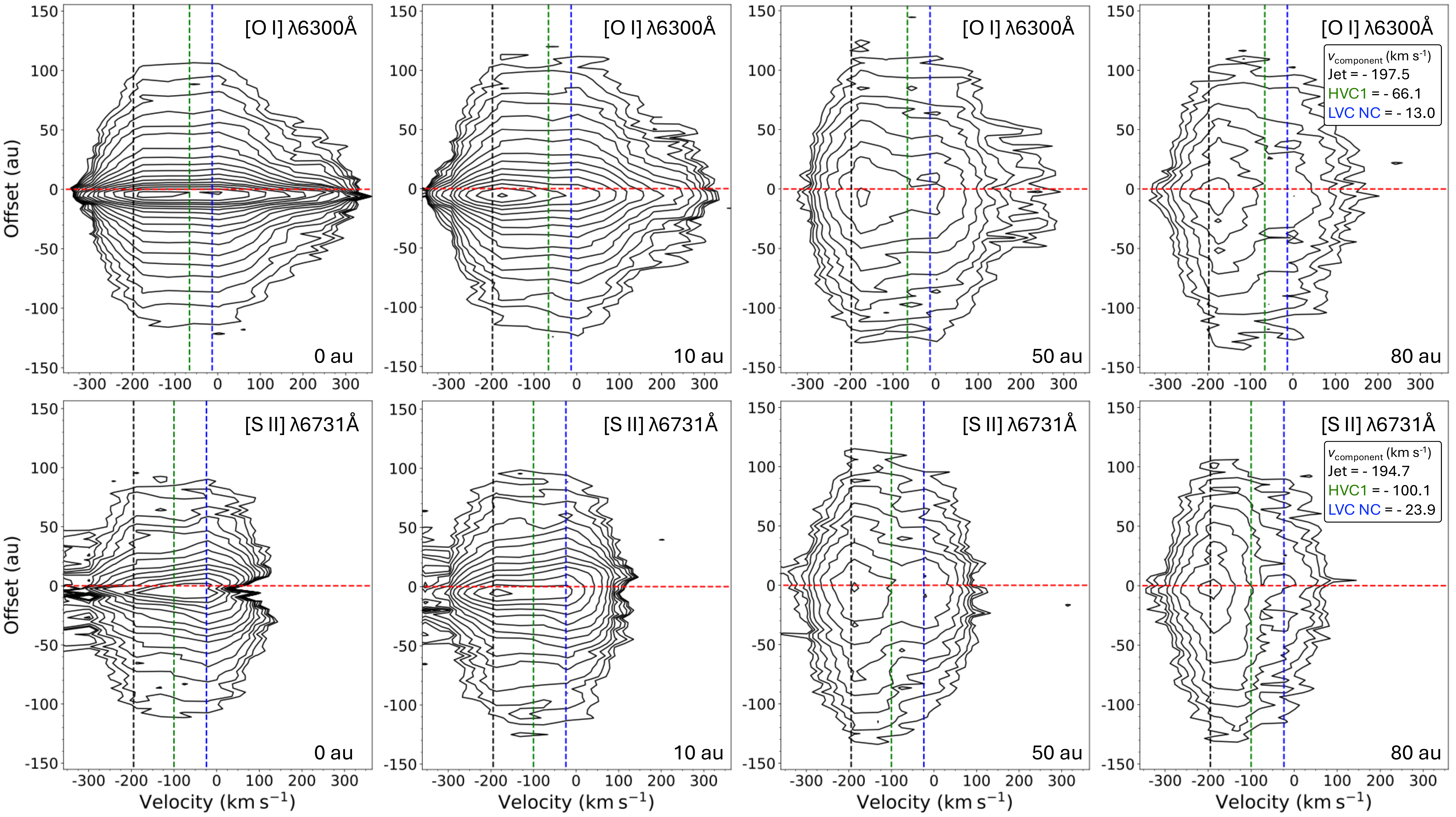}
   \caption{Position Velocity diagrams made transverse to the outflow axis for the \OIa\ and \SIIa\ emission line regions. Slices were extracted at the source position and at distances of 10~au, 50~au and 80~au from the star. Distances are not deprojected and these positions correspond to 29~au, 145~au and 232~au when corrected for the inclination of the system. The velocities of the jet, HVC1 and NC as measured by \cite{Whelan2021} are marked by the dashed lines. The NC appears as a clear second component distinct from the jet. The contour levels begin at 3$\sigma$ of the background emission and increase by factors of 1.5.}
        \label{PVacross}
    \end{figure*}

\begin{figure*}
   \centering
    \includegraphics[width=18cm, trim= 0cm 0cm 0cm 0cm, clip=true]{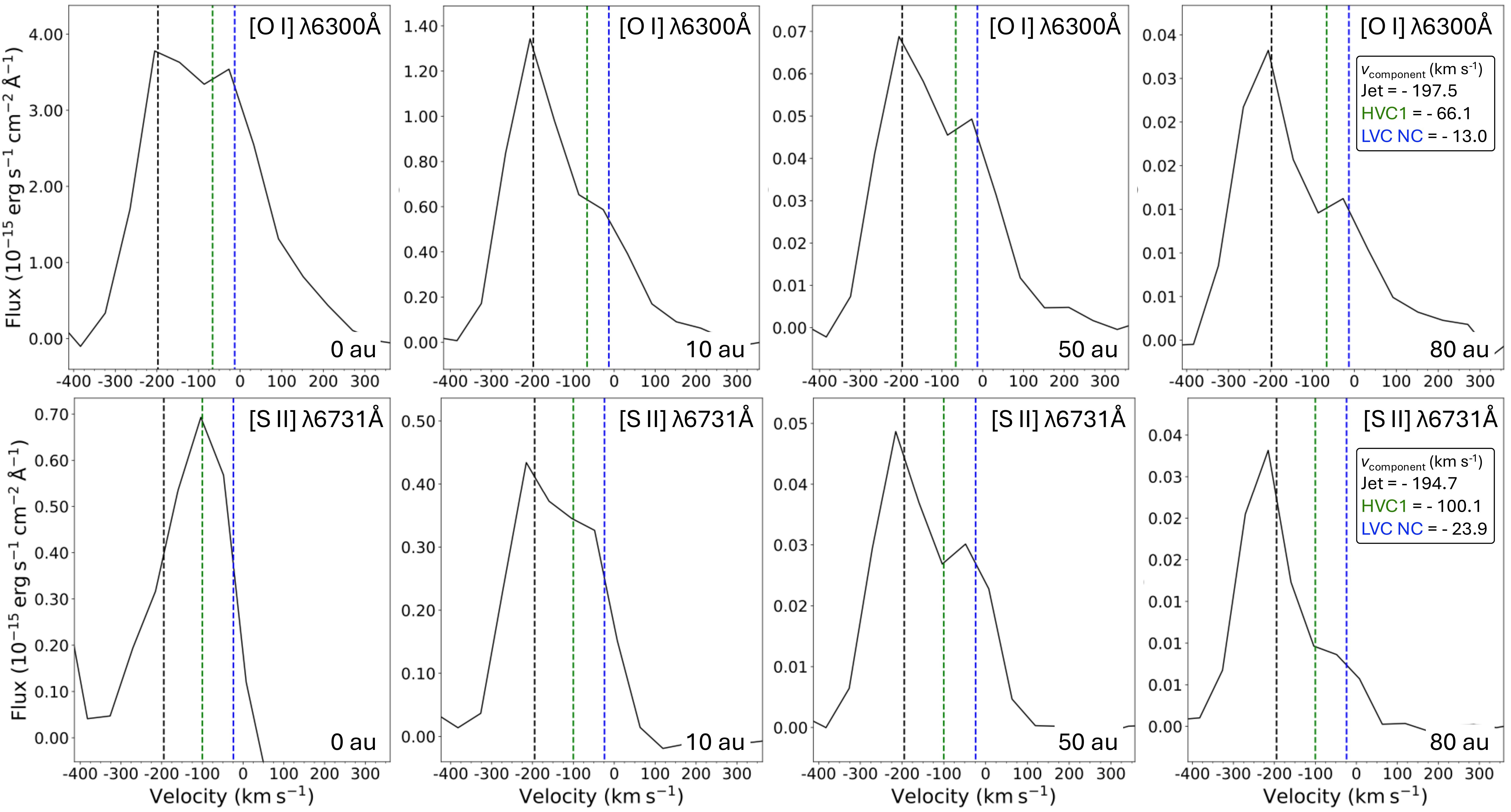}
   \caption{Line profiles extracted along the 0~au offset position (red dashed line) from the position velocity diagrams made transverse to the outflow axis, shown in Fig. \ref{PVacross}, for the \OIa\ and \SIIa\ emission line regions. These PV slices were extracted at the source position, and at distances of 10~au, 50~au and 80~au from the star. Distances are not deprojected and these positions correspond to 29~au, 145~au and 232~au when corrected for the inclination of the system. The velocities of the jet, HVC1 and NC as measured by \cite{Whelan2021} are marked by the dashed lines. The line flux decreases as the offset from the central source increases. The NC appears as a clear secondary component distinct from the jet.}
        \label{Fig: TransPVlineprofiles}
    \end{figure*}

Figure \ref{PVacross} presents position velocity (PV) diagrams, in \OIa\ and \SIIa\ emission, transverse to the outflow axis extracted at different distances along the outflow, namely, 0~au, 10~au, 50~au, and 80~au (0~au, 29~au, 145~au, and 232~au deprojected) from the source position. Initially, two velocity components are present; the jet, which is seen at all distances, and the NC, which fades as distance from the driving source of the outflow increases. This evidences that the NC is a separate outflow component to the jet. In order to better illustrate the presence of two distinct velocity components in the position velocity diagrams made transverse to the outflow axis, line profiles are extracted along the 0~au offset position, indicated by a horizontal red dashed line in Fig. \ref{PVacross}. In accordance with Fig. \ref{PVacross}, these line profiles are extracted from the \OIa\ and \SIIa\ emission line regions at distances of 0~au, 10~au, 50~au, and 80~au from the source position, and are presented in Fig. \ref{Fig: TransPVlineprofiles}. There are two distinct velocity components in the line profile, the jet and the NC. The NC is seen to decrease in intensity as the distance from the source increases. In contrast, the jet emission remains strong as the distance increases. The NC is not detected at the source position in the \SIIa\ line which is likely an effect of the lines critical density. One only begins to see the NC in \SIIa\ as the density decreases further from the driving source. \cite{Whelan2021} report no \SIIa\ emission in the NC within 2~au (8~au deprojected) of the source.
\end{multicols}

\FloatBarrier
\begin{multicols}{2}
\section{Aperture size for spectral extraction} \label{AppSec: ApertureSize}
\begin{figure*}[h!]
   \centering
    \includegraphics[width=18cm, trim= 0cm 0cm 0cm 0cm, clip=true]{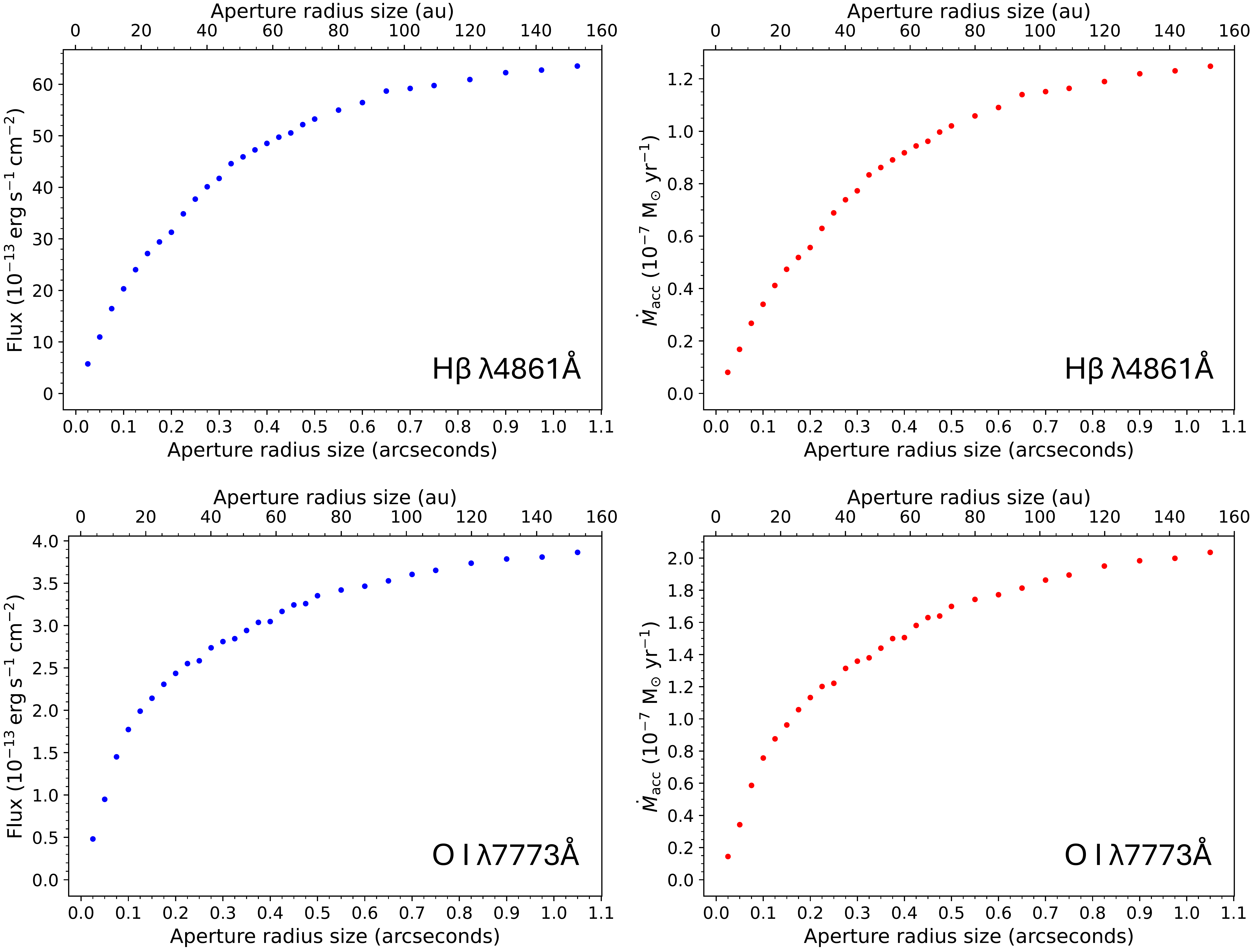}
   \caption{H$\beta$ $\lambda$4861 and \Oia$\,$flux (left panels) and $\dot{M}_{\rm acc}$ (right panels) as a function of aperture radius size. The aperture size in units of au is also displayed, calculated using the distance to RU~Lupi (d $\sim$ 140 pc).}
       \label{Fig: AppertureSize}
    \end{figure*}

Figure \ref{Fig: AppertureSize} displays how the measured flux and mass accretion rate ($\dot{M}_{\rm acc}$) increase as the radius of the spectral extraction aperture placed at the source position increases. This analysis was performed using the H$\beta$ $\lambda$4861 and \Oia$\,$emission lines to provide confirmation that this effect is present throughout the spectral axis. The angular resolution of the observation after adaptive optics is $\sim$ 0\farcs07, and the seeing is 0\farcs45. However, utilising the observations seeing as the aperture size, an aperture radius of $\sim$ 0\farcs23, for spectral extraction would result in a significant portion of the line flux not being accounted for, existing outside of the aperture radius. We instead chose to select an aperture size where both the line flux and consequently, the resulting mass accretion rate, begin to asymptote to a constant value. We choose an aperture radius of 0\farcs7 (98 au) to extract spectra from the source position. This is approximately three times the seeing of the observation. The results shown in Fig. \ref{Fig: AppertureSize} points to a large aperture for correction factor for MUSE NFM data, as could be expected due to MUSE's extended PSF wings. However, it is likely that the scattering discussed above is contributing to the flux at all wavelengths and that emission from the base of the outflow and accretion shocks would also be scattered, as well as contribution to emission from the disk surface, artificially inflating the measured line luminosities.
\end{multicols}

\FloatBarrier
\begin{multicols}{2}
\section{Mass accretion rate results}\label{AppSec: MassAccretionResults}
The accretion tracing emission lines are presented in Table \ref{Table: AccretionRates}. The mass accretion rates are derived following the method of \cite{Alcala2017}. The H$\alpha$ and [Ca II] lines were not used due to saturation at the source position. The mass accretion rate derived from the \OIa\ line was not used in the mean calculation, as we use this line to measure the mass outflow rates in the jet and NC. Notice the large uncertainties on $\dot{M}_{\rm acc}$ that are common when using line luminosities to estimate the mass accretion rate (See Sect. \ref{Sect: Discussion}). The estimated extinction for RU~Lupi is $\sim$ 0 \citep{Fang2018}.
\end{multicols}

\begin{table*}[h] 
\caption{Accretion luminosities and mass accretion rates in the different line tracers. H$\alpha$ and the [Ca\,{\scriptsize II}] triplet are saturated at the source position and thus could not be used as reliable accretion tracers. Furthermore, the mass accretion rate derived from the \OIa\ line was not used in the mean calculation, as we use this line to measure the mass outflow rates in the jet and NC.} 
\centering
\begin{tabular}{l c c c}
\hline
\hline
Line  &Log~$L_{\rm acc}$ (\Lsun) &Log~$\dot{M}_{\rm acc}$ (\Msun~yr$^{-1}$) &$\dot{M}_{\rm acc}$ (10$^{-7}$ \Msun~yr$^{-1}$) \\
\hline
H$\beta$ $\lambda$4861 & $-$ 0.15 $\pm$ 0.16 & $-$ 6.89 $\pm$ 0.16 & 1.28 $\pm$ 0.47 \\
\OIa  & $-$ 0.91 $\pm$ 0.41 & $-$ 7.66 $\pm$ 0.41 & 0.22 $\pm$ 0.21 \\
H$\alpha$ $\lambda$6563 & -  & - & -  \\
\Heib & $+$ 0.19 $\pm$ 0.33 & $-$ 6.55 $\pm$ 0.33 & 2.79 $\pm$ 2.12 \\
\Heic & $-$ 0.01 $\pm$ 0.29 & $-$ 6.75 $\pm$ 0.29 & 1.78 $\pm$ 1.19 \\
\Oia & $+$ 0.03 $\pm$ 0.49 & $-$ 6.71 $\pm$ 0.49 & 1.94 $\pm$ 2.19 \\
\Oib & $-$ 0.06 $\pm$ 0.62 & $-$ 6.81 $\pm$ 0.62 & 1.57 $\pm$ 2.24 \\
\CaIIa  & - & -  & - \\
\CaIIb  & -  & - & -  \\
\CaIIc  & -  & - & -  \\
Pa-10 $\lambda$9015 & $-$ 0.04 $\pm$ 0.43 & $-$ 6.78 $\pm$ 0.43 & 1.65 $\pm$ 1.64 \\
Pa-9 $\lambda$9230 & $-$ 1.08 $\pm$ 0.43 & $-$ 7.83 $\pm$ 0.43  & 0.15 $\pm$ 0.15\\
\hline 
\end{tabular}
\label{Table: AccretionRates}
\end{table*}

\FloatBarrier
\begin{multicols}{2}
\section{Outflow centroid positions} 
\begin{figure}[h!]
   \centering
  \includegraphics[width=8.6cm, trim= 0cm 0cm 0cm 0cm, clip=true]{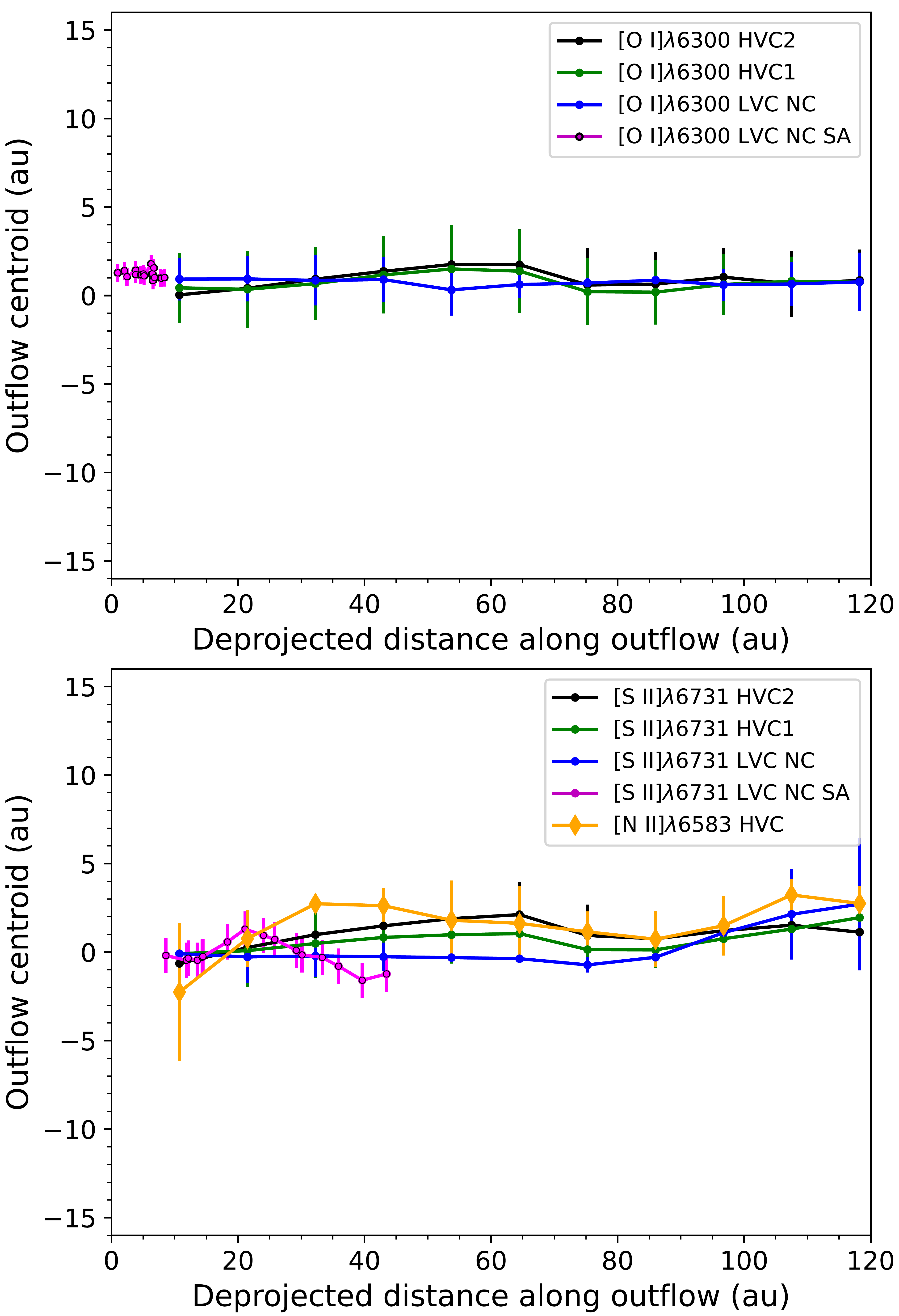}
    \caption{Outflow centroid measurements of the jet, HVC1 and NC in \OIa, \SIIa\ and \NII\ emission measured along the outflow with the driving source position at the origin. Centroid measurements as measured from the spectro-astrometry presented in \cite{Whelan2021} are over-plotted in magenta.}
    \label{Fig: centroids}
    \end{figure}

Outflow centroid measurements of the jet, HVC1 and NC in \OIa, \SIIa\ and \NII\ emission measured along the outflow with the driving source position at the origin are presented in Fig. \ref{Fig: centroids}. These outflow centroid are measured in the velocity component channel maps before deconvolution. The error bars on the data points are the difference in centroid position as measured in the pre-deconvolved channel maps and deconvolved channel maps. Centroid measurements as measured from the spectro-astrometry presented in \cite{Whelan2021} are over-plotted in magenta. The centroid measurements do not reveal any significant deviations from the position of the outflow axis, neither do they display any prominent outflow wiggling.
\end{multicols}

\FloatBarrier
\begin{multicols}{2}
\section{0104.C-0919(B) Observations} \label{AppSec: 0919(B) Obs}
\begin{figure*}[h!]
   \centering
      \includegraphics[width=18cm, trim= 0cm 0cm 0cm 0cm, clip=true]{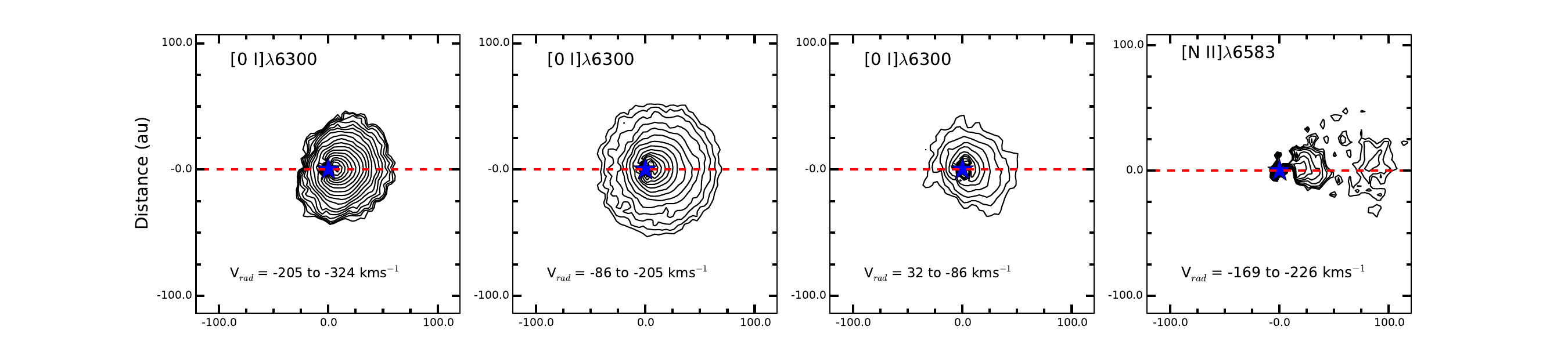}
\includegraphics[width=18cm, trim= 0cm 0cm 0cm 0cm, clip=true]{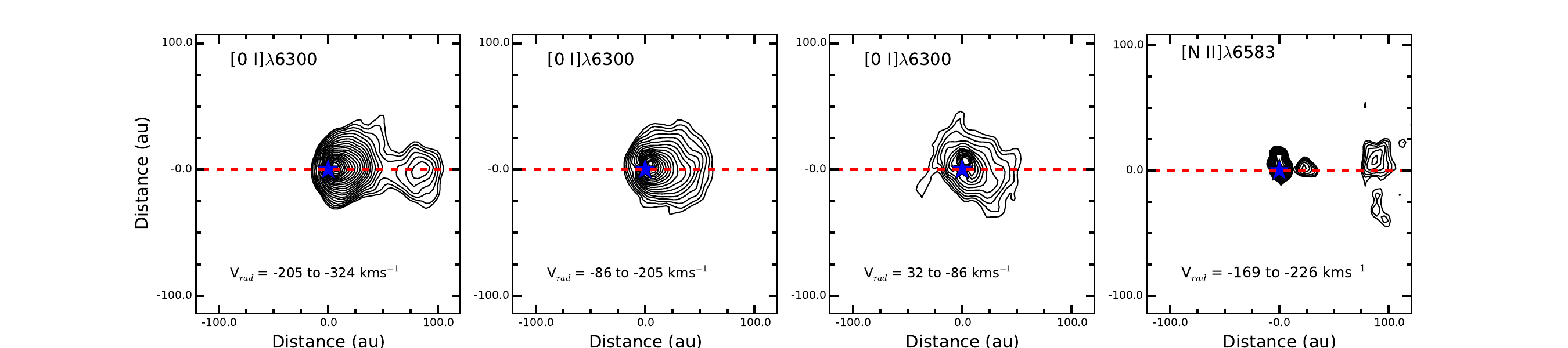}
   \caption{Comparing the channel maps produced using the 0104.C-0919(B) data of the jet, HVC1 and NC in \OIa\ emission, and the jet in \NII\ emission, before (top panels) and after (bottom panels) deconvolution. The contour levels begin at 3$\sigma$ of the background emission and increase by factors of 1.2.}
       %
       \label{Channelmaps_compOI}
    \end{figure*}

\begin{figure*}[h!]
   \centering
    \includegraphics[width=18cm, trim= 0cm 0cm 0cm 0cm, clip=true]{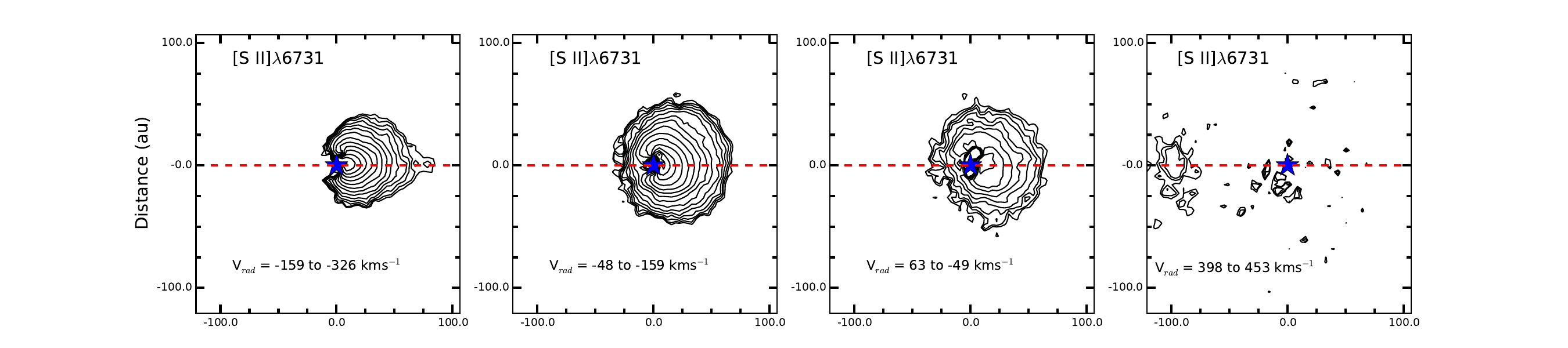}
  \includegraphics[width=18cm, trim= 0cm 0cm 0cm 0cm, clip=true]{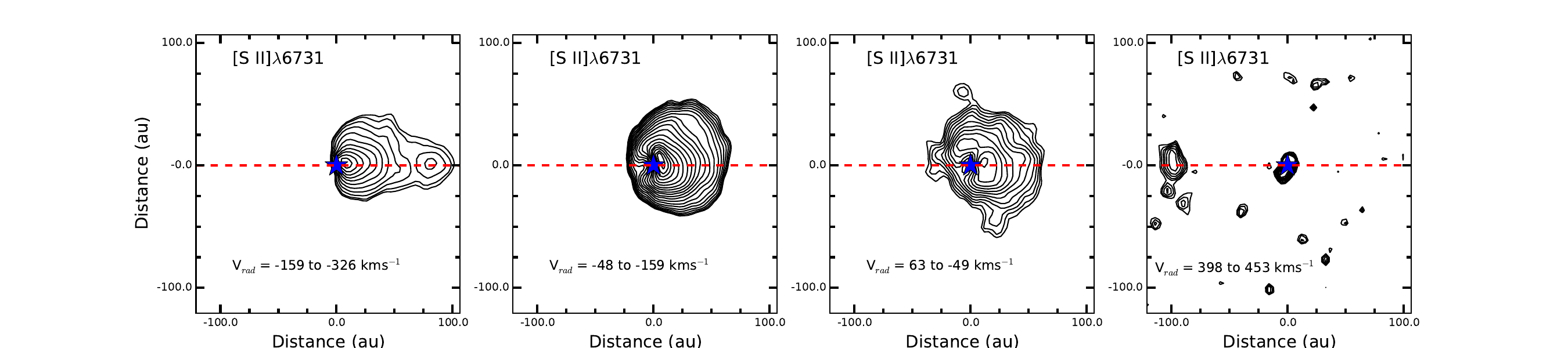}
   \caption{Comparing the channel maps produced using the 0104.C-0919(B) data of the jet, HVC1 and NC in \SIIa\ emission, and the high velocity redshifted knot, also in \SIIa\ emission, before (top panels) and after (bottom panels) deconvolution. The contour levels begin at 3$\sigma$ of the background emission and increase by factors of 1.2.}
       %
       \label{Channelmaps_compSII}
    \end{figure*}

RU~Lupi was also observed by MUSE in NFM in Summer 2021 as part of the programme 0104.C-0919 (B). These data were also analysed, and in Figs. \ref{Channelmaps_compOI} and \ref{Channelmaps_compSII}, we display the channel maps in \OIa, \SIIa\ and \NII\ emission before (top panels) and after (bottom panels) deconvolution prepared from this data. Although the morphological features identified in the channel maps are present here, comparing with Figs. \ref{Fig: OIchannels_compare} and \ref{Fig: SIIchannels_compare} shows the data to have a lower S/N. Additionally, these data were affected by a MUSE artefact known as a spike (MUSE manual: Section 3.9.1)\footnote{\url{https://www.eso.org/sci/facilities/paranal/instruments/muse/doc/ESO-261650_MUSE_User_Manual.pdf}}. This affected the \OIa\ channel maps significantly causing the measured PA to be different than the other outflow tracing lines, and for the emission to appear asymmetric. We therefore chose to make use of the 106.21EN.001 observations in our analysis.    
\end{multicols}

\end{appendix}
\end{document}